\documentclass[11pt]{article}
\usepackage[top=2cm,left=2cm,right=2cm]{geometry}
\usepackage[numbers,square]{natbib}

\usepackage[english]{babel}
\usepackage[latin1]{inputenc}
\usepackage{amssymb}
\usepackage{amsfonts}
\usepackage{amsmath}
\usepackage{fancyhdr}
\usepackage{epsfig}
\usepackage{graphics}
\usepackage{graphicx}
\usepackage{geometry}
\usepackage{subfigure}
\usepackage{rotating}
\usepackage{lineno}
\usepackage{pstricks}
\usepackage{color}

         \let\g = \gamma     \let\e = \epsilon

\newcommand{\nn}{\nonumber}

\newcommand{\beq}{\begin{equation}}
\newcommand{\eeq}{\end{equation}}
\newcommand{\bea}{\begin{eqnarray}}
\newcommand{\eea}{\end{eqnarray}}
\newcommand{\beqa}{\begin{eqnarray}}
\newcommand{\eeqa}{\end{eqnarray}}

\newcommand{\AL}[1]{\langle#1|}
\newcommand{\AR}[1]{|#1\rangle}

\newcommand{\SR}[1]{|#1]}
\renewcommand{\AA}[1]{\langle#1\rangle}
\newcommand{\SSS}[1]{[#1]}
\newcommand{\AS}[1]{\langle#1]}

\newcommand{\lid}[2]{#1\!\cdot\!#2}
\newcommand{\slashk}{k \! \! \!  /}

\newcommand{\slashp}{p \! \! \!  /}

\newcommand{\kapp}{\kappa}
\newcommand{\kapphat}{\hat{\kappa}}
\newcommand{\kstr}{\kappa^*}
\newcommand{\kstrhat}{\hat{\kappa}^*}
\newcommand{\qb}{\bar{q}}
\newcommand{\Amp}{\mathcal{A}}

\newcommand{\imag}{\mathrm{i}}

\newcommand{\graph}[3]{\raisebox{-#3ex}{\epsfig{file=figures/#1.pdf,width=#2ex}}}

\begin{document}

\begin{flushright}
IFJPAN-IV-2015-6
\end{flushright}

\begin{center}
\vspace{4.cm}
{\bf \large Amplitudes for High Energy Factorization via BCFW recursion}
\footnote{Presented at XXI Cracow Epiphany Conference on Future High Energy Colliders}

\vspace{1cm}

{\bf Mirko Serino}

\vspace{2mm}

\emph{E-mail}: mirko.serino@ifj.edu.pl

\vspace{0.5cm}

{The H. Niewodnicza\'nski Institute of Nuclear Physics \\
Polisch Academy of Sciences\\
Radzikowskiego 152, 31-342, Cracow, Poland\\}
\vspace{0.5cm}

\begin{abstract}

Recently, an extension of the BCFW on-shell recursion relation suitable to compute gauge invariant scattering amplitudes with off-shell
particles has been presented for Yang-Mills theories with fermions. 
In particular, 4- and 5-point amplitudes have been completely worked out.
The results are needed for the study of multi-parton scattering at hadron colliders in the framework
of High Energy Factorization (HEF).

\end{abstract}
\end{center}
%

\section{Introduction}

The computation of scattering amplitudes is an essential ingredient of high-energy physics.
The predictions of cross-sections at hadron colliders like the LHC are 
obtained by the use of factorization formulas, i.e. by convoluting the partonic scattering amplitudes, 
which describe the interaction of the elementary constituents of the colliding hadrons on a smaller time scale,
with universal functions (PDF's) describing the distributions of such partons inside the protons,
which account for evolution phenomena taking place on a time scale longer than the parton scattering itself.

The application of factorization formulas is justified by factorization theorems, which differ depending
on the kinematic regime in which they are meant to hold true.
In particular, in the {\em high-energy factorisation} or {\em $k_T$-factorization} approach~\cite{Catani:1990eg,Collins:1991ty,Catani:1994sq},
the amplitudes entering the calculation of cross sections feature particles with off-shell momenta, 
due to a non vanishing transverse component, which is additional w.r.t. the hadron (longitudinal) 
momentum fraction carried by the parton.

In order for results to be physical, amplitudes need to be gauge invariant, 
a property whose definition is far form trivial in case there are off-shell legs. 
In most of the approaches existing in the literature on scattering amplitudes with off-shell gluons~\cite{Lipatov:1995pn,Antonov:2004hh,Kotko:2014aba} 
Wilson lines are a basic ingredient of the procedure. 
A different approach to the problem was presented in \cite{vanHameren:2012if,vanHameren:2013csa},
where a series of eikonal Feynman rules are derived which allow to draw, beside the ordinary Feynman diagrams,
gauge-restoring contributions which eventually yield the same results already known in the literature. 
While the inclusion of fermions has so far eluded the approach in terms of Wilson lines, the diagrammatic approach
is hardly suitable for amplitudes containing more then 4 or 5 partons, as the number of diagrams to be computed
becomes overwhelming already at tree level, just as in the on-shell kinematics.

A dramatic improvement in the calculation of scattering amplitudes has been achieved ever since 2005 after the
introduction of the BCFW recursion procedure, originally presented for pure Yang-Mills theories~\cite{Britto:2004ap,Britto:2005fq}
and later extended to include amplitudes with fermions \cite{Luo:2005rx}.
The question whether this recursion can be generalized to amplitudes with off-shell partons was tackled 
for the first time in \cite{vanHameren:2014iua} in the case of gluons,leading to the release of a code for the numerical
evaluation of such amplitudes~\cite{Bury:2015dla}. 
The procedure was extended to amplitudes with a fermion pair in \cite{vanHameren:2015bba}.

\section{Definitions}\label{definitions}%

We always consider scattering amplitudes with all particles outgoing.
The momentum $k^\mu$ can be decomposed in terms of its light-like direction $p^\mu$, 
satisfying $\lid{p}{k}=0$ and, if the particle is off-shell, of a transversal part, following
%
\begin{equation}
k^\mu = x(q)p^\mu - \frac{\kapp}{2}\,\frac{\AS{p|\gamma^\mu|q}}{\SSS{pq}}  - \frac{\kstr}{2}\,\frac{\AS{q|\gamma^\mu|p}}{\AA{qp}} \, ,
\end{equation}
%
with $q^\mu$ an auxiliary lightlike 4-momentum
%
\begin{equation}
x(q)=\frac{\lid{q}{k}}{\lid{q}{p}}
\quad,\quad
\kapp = \frac{\AS{q|\slashk|p}}{\AA{qp}}
\quad,\quad
\kstr = \frac{\AS{p|\slashk|q}}{\SSS{pq}} \, .
\end{equation}
The coefficients $\kapp$ and $\kstr$ can be shown to be 
independent of the auxiliary momentum $q^\mu$, in the sense that any other
lightlike vector $q'$ can be used in its place, provided $\lid{k}{q'} \neq 0$ and
\begin{equation}
k^2 = -\kapp\kstr \, .
\end{equation}
We consider  {\em color-ordered\/} or {\em dual\/} amplitudes, 
which contain only planar Feynman graphs and are constructed with color-stripped Feynman rules.

In the diagrams in the eikonal quarks will be denoted by dashed solid lines.
If we want to specify that an external gluon or fermion is off-shell,
we represent it with a double curly line and a double straight line respectively.
Notice that one can imagine that the double lines representing off shell partons can be bent apart:
in the case of an off-shell gluon, they will thus form a single eikonal quark line;
in the case of an off-shell fermion, they will split into an eikonal quark line and an auxiliary null momentum photon ($\gamma_A$).

The Feynman rules allowing to compute gauge-invariant scattering amplitudes
for processes with off-shell gluons and fermions were worked out in 
\cite{vanHameren:2012if,vanHameren:2013csa} respectively and are explicitly reported in \cite{vanHameren:2015bba}.

It is decisive to stress that the BCFW program is completely independent of the diagrammatic approach. 
In fact, also 3-point functions with off-shell particles can be found by applying
the recursion itself, via the $C$ and $D$ residues to be discussed in the following.
These, in turn, require only the knowledge of on-shell $3$-point functions,
which are completely fixed by symmetry requirements.
Once one knows $3$-point functions, all higher-point amplitudes follow naturally.

\subsection{The 3-point amplitudes}\label{App3Point}%

In this short section, we list the $3$-point amplitudes which are the very building blocks of the BCFW recursion.

The on-shell 3-point amplitudes, which are well known from the literature,
vanish for on-shell real momenta, but not when some of these become complex.
The same is true for on-shell 3-point amplitudes with one fermion pair and one gluon, which can be
seen from the general MHV representation in \cite{Luo:2005rx}.
Here they are,
\begin{eqnarray}
\Amp(g_1^+,g_2^-,g_3^-) = \frac{\AA{23}^3}{\AA{12}\AA{31}} \, 
\quad
&&
\Amp(g_1^-,g_2^+,g_3^+) = \frac{\SSS{32}^3}{\SSS{21}\SSS{13}} \nn
\\
\Amp(g^-,\qb^+,q^-) = \frac{\AA{g q}^3\AA{g \qb}}{\AA{g \qb}\AA{\qb q}\AA{q g}}\, 
\quad
&&
\Amp(g^+,\qb^+,q^-) = \frac{\SSS{g \qb}^3\SSS{g q}}{\SSS{g q}\SSS{q \qb}\SSS{\qb g}} \nn
\\
\Amp(g^-,\qb^-,q^+) = \frac{\AA{g \qb}^3\AA{g q}}{\AA{g \qb}\AA{\qb q}\AA{q g}}
\quad
&&
\Amp(g^+,\qb^-,q^+) = \frac{\SSS{g q}^3\SSS{g \qb}}{\SSS{g q}\SSS{q\qb}\SSS{\qb g}} \, .
\end{eqnarray}
\\

Amplitudes with three gluons, one of which is off-shell, were computed in \cite{vanHameren:2014iua}.
Equal helicity amplitudes vanish,
\begin{equation}
\Amp(g_1^*,g_2^+,g_3^+) = \Amp(g_1^*,g_2^-,g_3^-) = 0 \, . 
\end{equation}
Amplitudes for which the two on-shell gluons have opposite helicity are given by a simple generalization of the on-shell formula,
\begin{equation}
\Amp(g_1^*,g_2^+,g_3^-) = \frac{1}{\kstr_1}\frac{\AA{31}^3}{\AA{12}\AA{23}} = \frac{1}{\kappa_1}\frac{\SSS{21}^3}{\SSS{13}\SSS{32}}   \, .
\end{equation}
\\

The 3-point amplitudes with one on-shell fermion pair and  an off-shell gluon have \emph{two equivalent representations},
\begin{eqnarray}
\Amp(g^*,\qb^+,q^-)  &=& 
\frac{1}{\kappa_g}\, \frac{\SSS{g \qb}^3 \SSS{g q} }{\SSS{g q} \SSS{q \bar{q} } \SSS{\bar{q}  g} }  = 
\frac{1}{\kappa^*_g}\, \frac{\AA{g q}^3 \AA{g \qb} }{\AA{g \bar{q}  } \AA{ \bar{q} q } \AA{q g}   } 
\nn \\
\Amp(g^*,\qb^-,q^+)  &=&
\frac{1}{\kappa_g}\, \frac{\SSS{g q}^3 \SSS{g \qb} }{\SSS{g q} \SSS{q \bar{q} } \SSS{\bar{q}  g} }  = 
\frac{1}{\kappa^*_g}\, \frac{\AA{g \qb}^3 \AA{g q} }{\AA{g \bar{q}  } \AA{ \bar{q} q } \AA{q g}   }   \, .
\end{eqnarray}
\\

The 3-point amplitudes with one off-shell antifermion and off-shell fermion respectively are
\begin{eqnarray}
\Amp(g^+,\qb^*,q^-) &=&  \frac{1}{\kappa_{\qb}}\, \frac{ \SSS{g \qb}^3 \SSS{g q} }{ \SSS{g q} \SSS{q \qb} \SSS{\qb g} }  \, ,
\nn \\
\Amp(g^-,\qb^*,q^+) &=&  \frac{1}{\kstr_{\qb}}\, \frac{ \AA{g \qb}^3 \AA{g q} }{\AA{g \qb} \AA{\qb q} \AA{q g} } \, ,
\nn \\
\Amp(g^\pm,\qb^*,q^\pm) &=& 0 \, .
\end{eqnarray}
and
\begin{eqnarray}
\Amp(g^+,\qb^-,q^*) &=&  \frac{1}{\kappa_{q}}\, \frac{ \SSS{g q}^3 \SSS{g \qb} }{ \SSS{g q} \SSS{q \qb} \SSS{\qb g} }  \, ,
\nn \\
\Amp(g^-,\qb^+,q^*) &=&  \frac{1}{\kstr_{q}}\, \frac{ \AA{g q}^3 \AA{g \qb} }{\AA{g \qb} \AA{\qb q} \AA{q g} } \, ,
\nn \\
\Amp(g^\pm,\qb^\pm,q^*) &=& 0 \, .
\end{eqnarray}
%

\section{BCFW recursion}\label{recursion}%

For every particle with momentum $k^\mu_i$, an orthogonal direction $p^\mu_i$ is given,
\begin{align}
k_1^\mu + k_2^\mu + \cdots + k_n^\mu = 0
&\qquad\textrm{momentum conservation}\label{Eq:momcons}\\
p_1^2 = p_2^2 = \cdots = p_n^2 = 0
&\qquad\textrm{light-likeness}\label{momcon1}\\
\lid{p_1}{k_1} = \lid{p_2}{k_2}=\cdots=\lid{p_n}{k_n}=0 \, .
&\qquad\textrm{eikonal condition}
\label{Eq:eikcon}
\end{align}
In the case of an on-shell particle, direction and momentum are the same vector.

The polarization vectors for gluons can be expressed as
\begin{equation} \label{polarization}
\e^\mu_{+ } = \frac{\AS{q | \g^\mu | g}}{\sqrt{2}\AA{q g}} \, , 
\quad
\e^\mu_{-} = \frac{\AS{g |\g^\mu | q}}{\sqrt{2}\SSS{g q}} \, ,
\end{equation}
where $q$ is the auxiliary lightlike vector and $g$ is a short-hand notation for the gluon momentum.
We denote gluon spinors by the numbers of the corresponding particles, 
whereas quarks and antiquarks are always indicated by $q$ and $\qb$ respectively.

The starting point of the on-shell BCFW recursion relation is the residue theorem
\begin{equation}
\lim_{z \to \infty} f(z) = 0 \Rightarrow \oint \frac{dz}{2\pi\imag} \, \frac{f(z)}{z} = 0 \, ,
\end{equation}
where the integration countour encloses all the poles of the rational function $f(z)$ and extends to infinity, implying that 
the function at the origin $f(0)$ is given by the sum over the residues at the single poles in the complex plane,
\begin{equation}
f(0) = - \sum_{i} \frac{\lim_{z\rightarrow z_i} f(z)\, (z-z_i) }{z_i} \, .
\label{residues}
\end{equation}
Now, if $f(z)=\Amp(z)$, where $\Amp(z)$ is a scattering amplitude which is turned into 
a function of a complex variable without spoiling momentum conservation and on-shellness, 
it can be shown that it is enough to identify the single poles in $z$ appearing in some of the propagators
in order to reconstruct the amplitude in terms of simpler building blocks. 
These are found to be products of on-shell lower-point amplitudes 
times an intermediate propagator, on the ground of general unitarity requirements~\cite{Britto:2004ap,Britto:2005fq}.

In order to make a scattering amplitude a rational function of a complex variable $z$
in a way that suits the off-shel case, two particles are picked up, say $i$ and $j$, and each particles's direction is chosen to
be the reference vector for the other, so that their momenta with transverse component are
\begin{eqnarray}
k_i^\mu = x_i(p_j)\, p_i^\mu - \frac{\kappa_i}{2}\, \frac{\AL{i} \g^\mu \SR{j}}{\SSS{ij}} - \frac{\kstr_i}{2}\, \frac{\AL{j} \g^\mu \SR{i}}{\AA{ji}}
\nn \\
k_j^\mu = x_j(p_i)\, p_j^\mu - \frac{\kappa_j}{2}\, \frac{\AL{j} \g^\mu \SR{i}}{\SSS{ji}} - \frac{\kstr_j}{2}\, \frac{\AL{i} \g^\mu \SR{j}}{\AA{ij}} \, .
\end{eqnarray}
Let the shift vector be
\beq
e^\mu = \frac{1}{2}\, \AL{i} \g^\mu \SR{j} \, , \quad  
p_i \cdot e = p_j \cdot e = e \cdot e = 0 \, .
\eeq
The shifted (hatted) momenta are thus written as
\bea
\hat{k}_i^\mu = k_i + z e^\mu 
&=&
x_i(p_j)\, p_i^\mu - \frac{\kappa_i - z\SSS{ij}}{2}\, \frac{\AL{i} \g^\mu \SR{j}}{\SSS{ij}} - \frac{\kstr_i}{2}\, \frac{\AL{j} \g^\mu \SR{i}}{\AA{ji}}
\nn \\
&=& 
x_i(p_j)\, p_i^\mu - \frac{\hat{\kappa}_i}{2}\, \frac{\AL{i} \g^\mu \SR{j}}{\SSS{ij}} - \frac{\kstr_i}{2}\, \frac{\AL{j} \g^\mu \SR{i}}{\AA{ji}}
\nn \\
\hat{k}_j^\mu = k_j - z e^\mu 
&=&
x_j(p_i)\, p_j^\mu - \frac{\kappa_j}{2}\, \frac{\AL{j} \g^\mu \SR{i}}{\SSS{ji}} - \frac{\kstr_j + z \AA{ij}}{2}\, \frac{\AL{i} \g^\mu \SR{j}}{\AA{ij}}
\nn \\
&=&
x_j(p_i)\, p_j^\mu - \frac{\kappa_j}{2}\, \frac{\AL{j} \g^\mu \SR{i}}{\SSS{ji}} - \frac{\kstrhat_j}{2}\, \frac{\AL{i} \g^\mu \SR{j}}{\AA{ij}} \, .
\label{shifts}
\eea
Momentum conservation and either on-shellness or the eikonal conditions 
$p_i \cdot \hat{k}_i = 0$ and $p_j \cdot \hat{k}_j = 0$ are preserved by the shift (\ref{shifts}).
If we had chosen $e^\mu = 1/2\, \AL{j}\g^\mu\SR{i}$ the shifted quantities would have been $\kstrhat_i$ and $\kappa_j$.
The changes induced either in the momenta or in the directions by the two possible shift vectors can be concisely summarized as follows:
\begin{eqnarray}
e^\mu &=& \frac{1}{2}\AL{i}\g^\mu\SR{j} 
\Leftrightarrow 
\left\{ \begin{array}{c}
\textrm{$i$ off-shell:} \quad \kapphat_i = \kappa_i - z \SSS{ij} 
\\
\textrm{$i$ on-shell:} \quad \SR{\hat{i}} = \SR{i} + z \SR{j} 
\\
\textrm{$j$ off-shell:} \quad \kstrhat_i = \kstr_j + z \AA{ij} 
\\
\textrm{$j$ on-shell:} \quad \AR{\hat{j}} = \AR{j} - z \AR{i}
\end{array} \right.
\label{shift1}
\\
e^\mu &=& \frac{1}{2}\AL{j}\g^\mu\SR{i}
\Leftrightarrow 
\left\{ \begin{array}{c}
\textrm{$i$ off-shell:} \quad \kapphat_i = \kstr_i - z \AA{ji} 
\\
\textrm{$i$ on-shell:} \quad \AR{\hat{i}} = \AR{i} + z \AR{j} 
\\
\textrm{$j$ off-shell:} \quad \kapphat_j = \kappa_j + z \SSS{ji} 
\\
\textrm{$j$ on-shell:} \quad \SR{\hat{j}} = \SR{j} - z \SR{i}
\end{array} \right.
\label{shift2}
\end{eqnarray}
It is basic to the BCFW argument that (\ref{shift1}) and (\ref{shift2})
imply that the large $z$ behaviours of the polarization vectors of shifted gluons are
\begin{eqnarray}
e^\mu &=& \frac{1}{2}\AL{i}\g^\mu\SR{j} \Rightarrow \epsilon^\mu_{i-} \sim \frac{1}{z} \quad \textrm{and} \quad \epsilon^\mu_{j+} \sim \frac{1}{z} \, ,
\nn \\
e^\mu &=& \frac{1}{2}\AL{j}\g^\mu\SR{i} \Rightarrow \epsilon^\mu_{i+} \sim \frac{1}{z} \quad \textrm{and} \quad \epsilon^\mu_{j-} \sim \frac{1}{z} \, ,
\label{shift3}
\end{eqnarray}
whereas the opposite helicity polarization vectors of shifted gluons stay constant. 
It is basic for us, in order for our argument to work, to include in our amplitudes the propagators of the
external off-shell particles, who play the same role as the gluon  polarization vectors in the on-shell case.

Not all of the shift vector choices are are suitable to apply the BCFW recursion, 
because some of them violate the basic hypothesis of the residue theorem
\begin{equation}
\lim_{z\to\infty} \Amp(z) = 0 \, .
\end{equation}

In \cite{Britto:2005fq} it was found that with the shift vector $e^\mu = \frac{1}{2} \AL{i}\g^\mu\SR{j}$ 
$\Amp(z) \stackrel{z\to\infty}{\longrightarrow} 0$  for three possible helicity choices of the shifted particles, namely $(h_i,h_j) = (-,+),(-,-),(+,+)$.
It is easy to obtain a diagrammatic proof for the first case, that we dub the \emph{original BCF prescription}, 
and all of our results for our amplitudes with $1$ off-shell particle refer to shifts which reduce to this case or, 
if $e^\mu=\frac{1}{2}\AL{j}\g^\mu\SR{i}$, to $(h_i,h_j) = (+,-)$. 

As eikonal quark vertices only depend on the direction and eikonal quark propagators can only contribute powers of $z$ to the denominator,
the BCFW proof extends to our case without much trouble, if we decide to shift only on-shell gluons.
This also works for amplitudes with a fermion pair~\cite{Luo:2005rx}.
Then, if one of the shifted gluons is off-shell and the other one is on-shell we require the helicity 
of the latter to agree with the original BCF prescription~\cite{vanHameren:2014iua}.
Finally, if both shifted gluons are off-shell, they both will contribute a factor $1/z$ with either shift vector choice.
It is quite general that \emph{if both shifted particles are off-shell, BCFW recursion works with both shift vectors}. \\

One may ask what happens when shifting one gluon and one fermion line.
In~\cite{Luo:2005rx} it was stressed that for on-shell amplitudes 
it is not allowed to shift one fermion and one gluon with the same helicity sign.
So in our case we can choose to shift an off-shell (anti) fermion and an on-shell gluon
provided that, in the on-shell limit, the (anti) fermion has opposite helicity sign w.r.t.\ the gluon.

There is another, more peculiar possibility, typical of the fermion case: 
we can choose the shift vector in such a way that the shifted particles will have the same helicity sign in the on-shell limit, 
provided the gluon polarization behaves like $1/z$: this is because, as long as the fermion is off-shell, its direction
does not shift, whereas its transverse momentum does, and this ensures that the amplitude will have the correct
asymptotic behaviour; in on-shell limit this would not work. In a sense, off-shell amplitudes are better behaved.\\

\subsection{The residues}\label{residuesec}%

The single poles in $z$ appear due to the denominators of the gluon or fermion propagators.
Our scattering amplitude $\Amp(0)$ is given by
\begin{equation}
\Amp(0) = \sum_{s=g,f} \left(  \sum_{p} \sum_{h=+,-} \mathrm{A}^s_{p,h} + \sum_{i} \mathrm{B}^s_i  + \mathrm{C}^s + \mathrm{D}^s \right) \, .
\end{equation}
The index $s$ refers to the particle species, namely gluons or fermions; $h$ is the helicity;
$K^\mu$ denotes the momentum flowing through the propagators exhibiting poles. 
The residues in the case of gluon poles are

\begin{align}
\mathrm{A}^g_{p,h} &= \graph{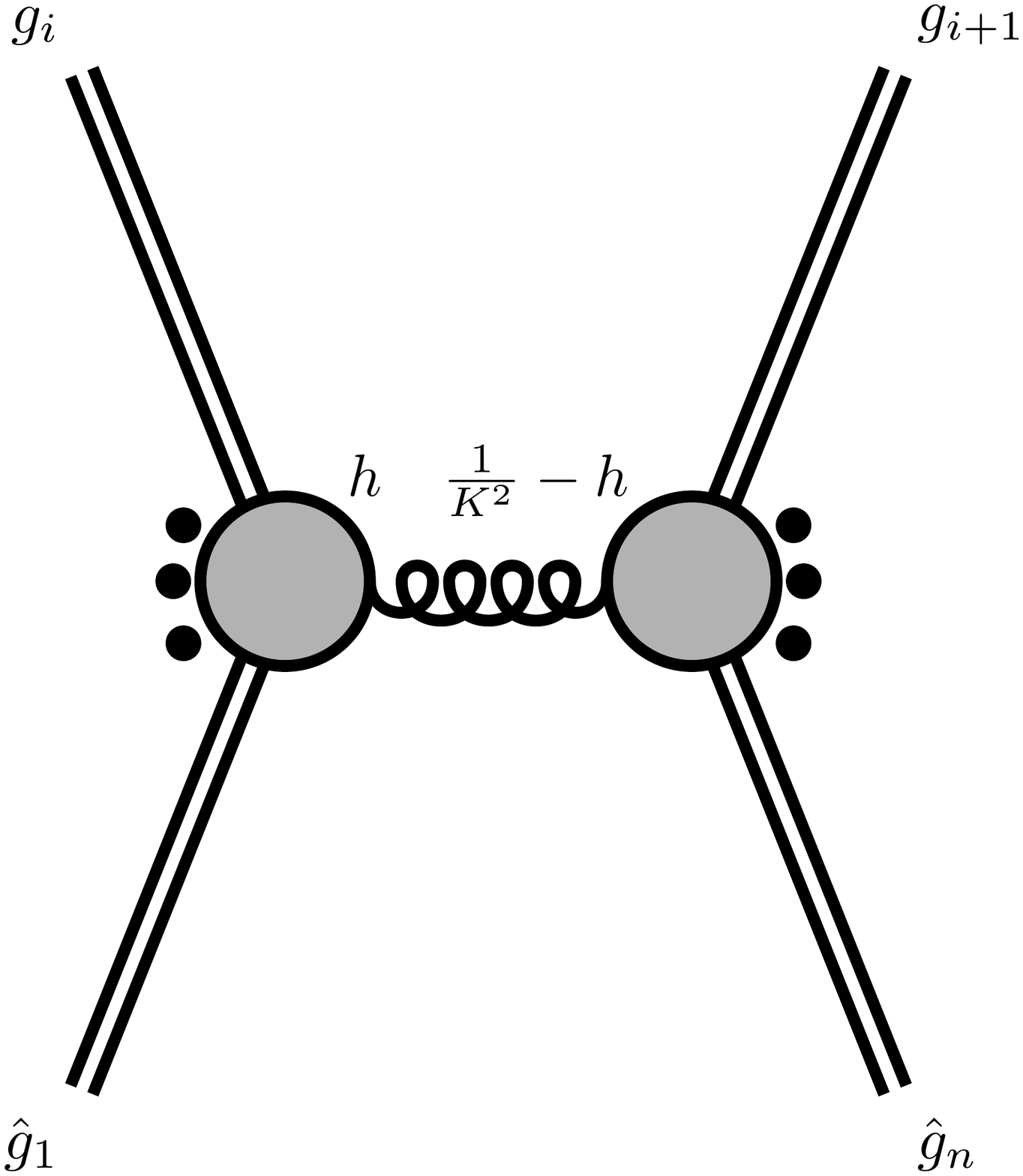}{22}{15}
\quad&
\mathrm{B}^g_{i} &= \graph{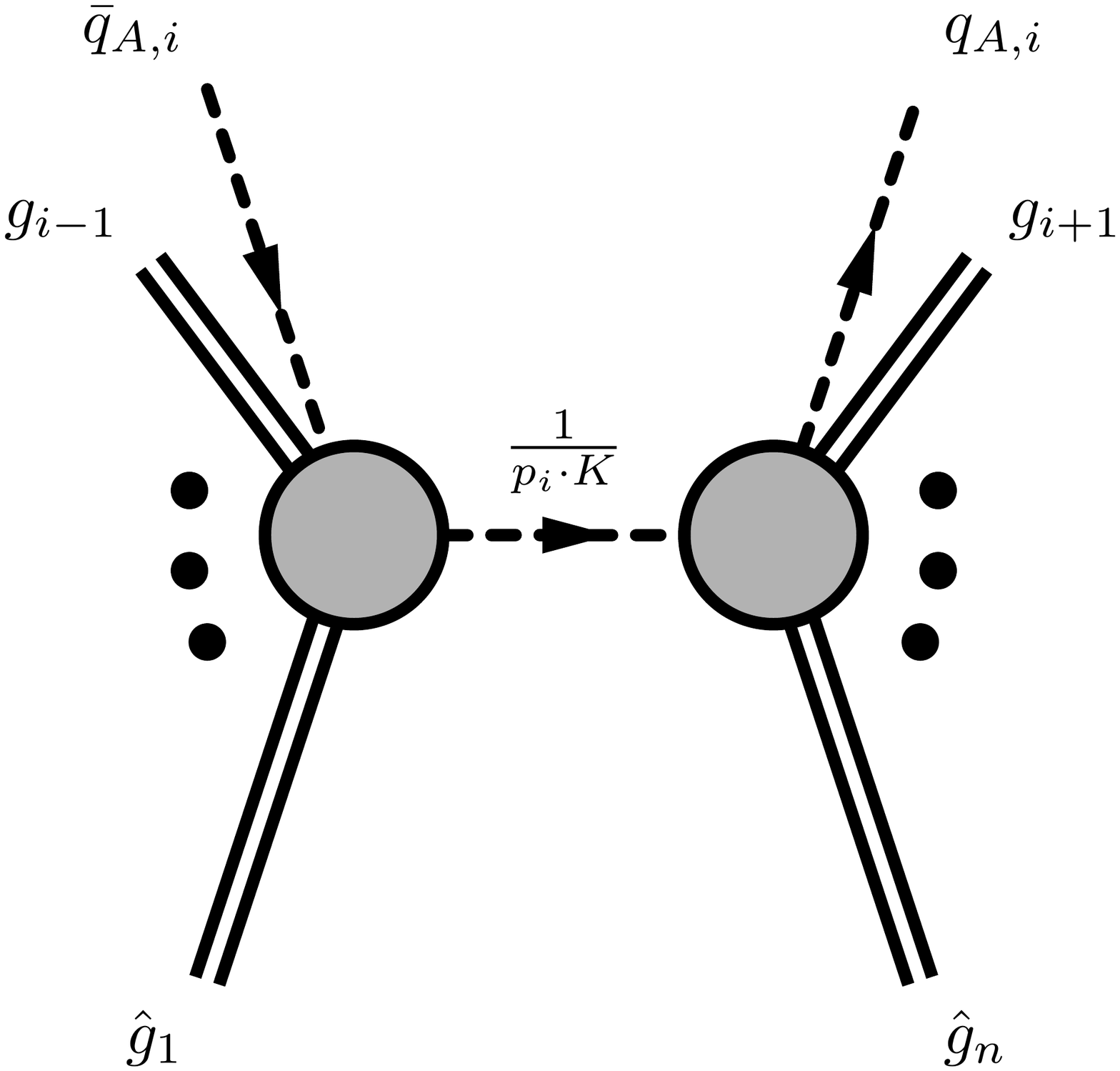}{22}{15}
\notag\\
\mathrm{C}^g &= \graph{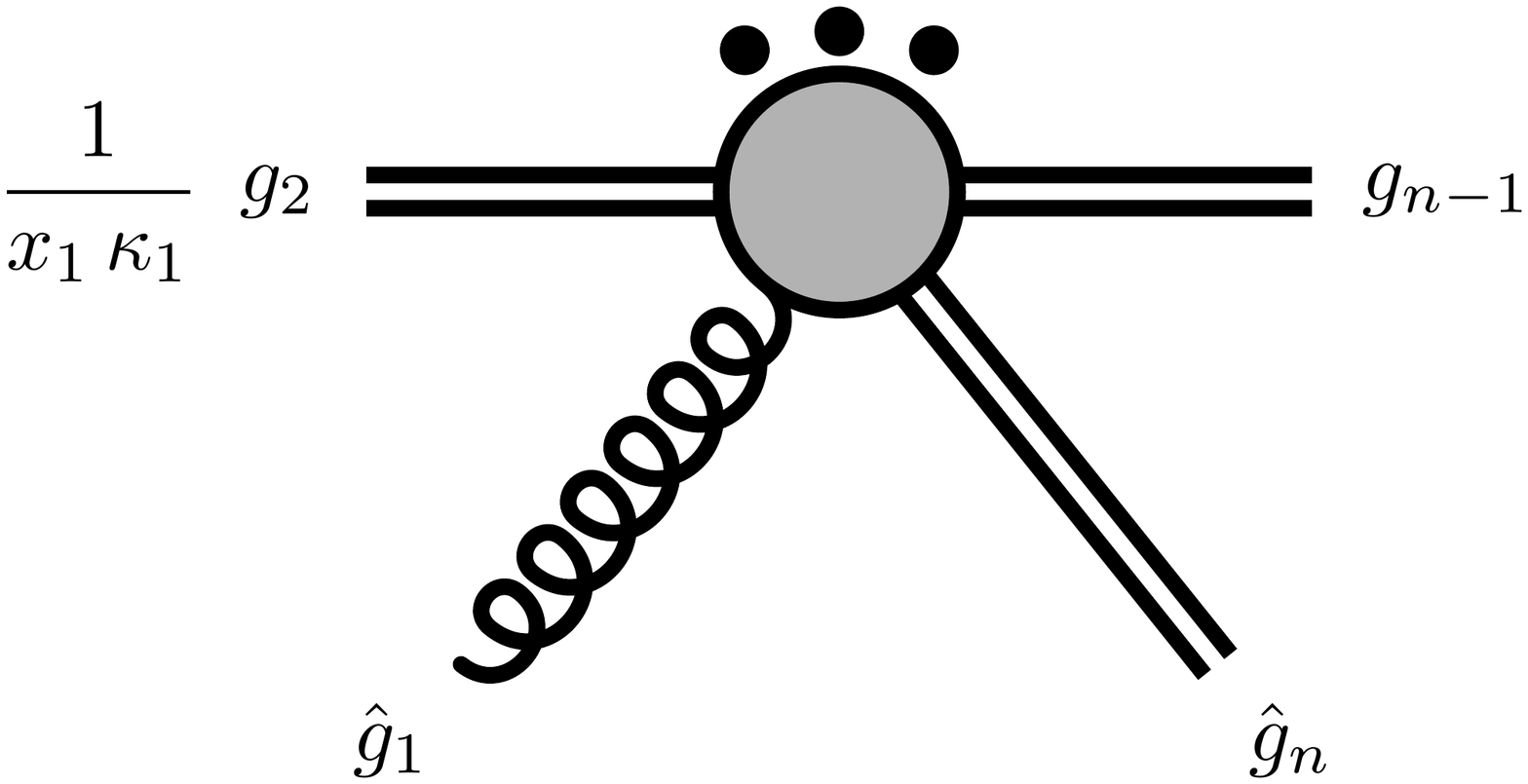}{25}{20}
\quad&
\mathrm{D}^g &= \graph{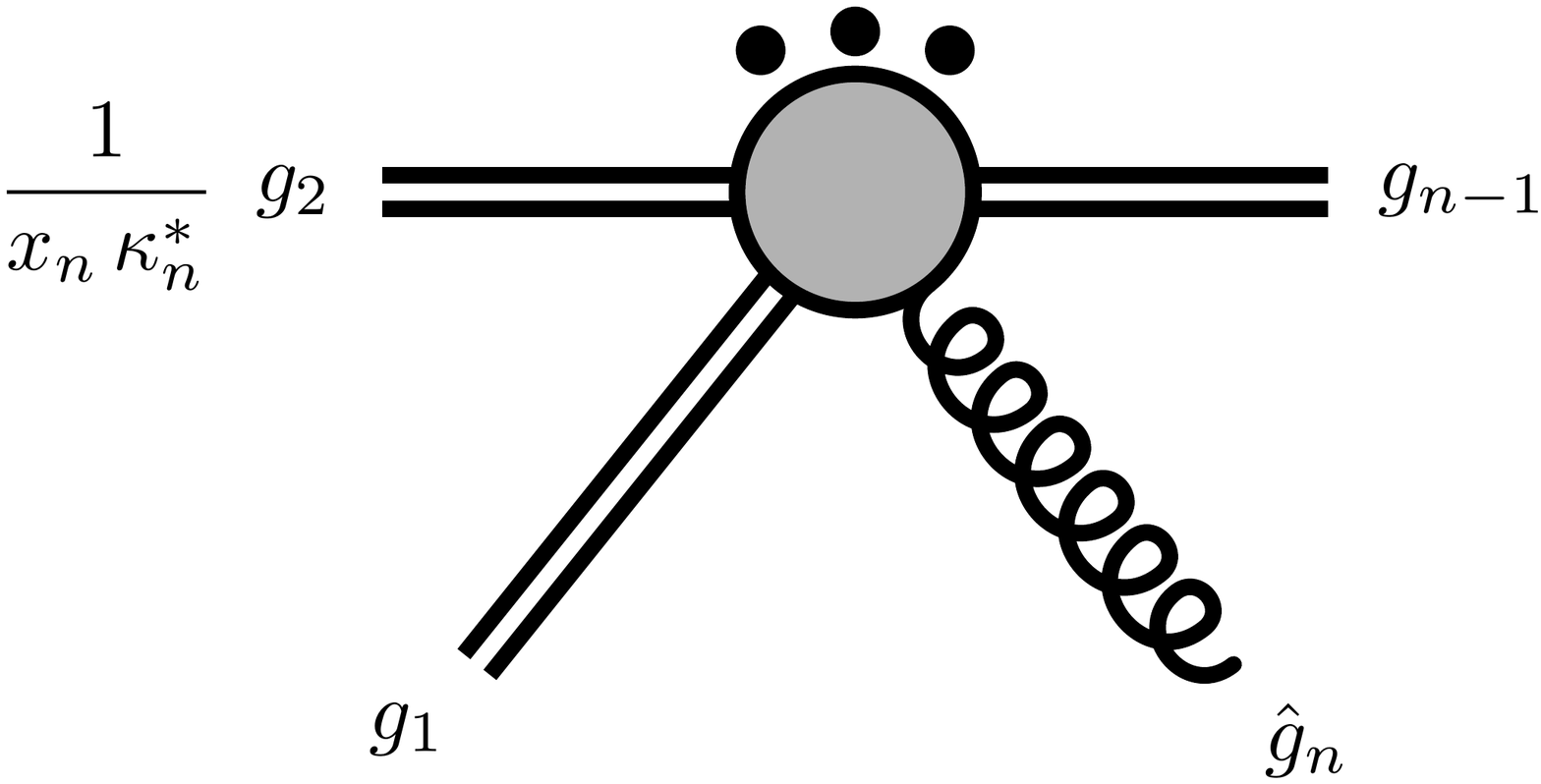}{25}{20}
\nn
\end{align}

%
$\mathrm{A}^g_{p,h}$ are due to the poles which appear in the original BCFW recursion. 
The index $p$ stands for the cyclically ordered distributions of the particles into two subsets; the shifted particles
are never on the same sub-amplitude. The pole is due to an intermediate virtual gluons, 
whose shifted momentum squared is $K^2(z)$ is on-shell for $$ z = - \frac{K^2}{2 \, e \cdot K} \, . $$

$\mathrm{B}^g_i$ residues are due to poles appearing in the auxiliary eikonal quarks propagators whose denominator vanishes.
This means $p_i \cdot \hat{K}(z) = 0$, where $\hat{K}$ is the momentum flowing through the propagator. 
The location of these pole is $$  z = - \frac{2\, p_i \cdot K}{2\,p_i \cdot e}\, . $$ If the $i$-th particle is on-shell, this term is not present.
         
$\mathrm{C}^g$ and $\mathrm{D}^g$ denote the same kind of residues: they appear respectively when the shifted $i$-th or $j$-th gluons are off-shell.
They are due to the vanishing of the shifted momentum in the propagator: $k_i^2(z) = 0 \quad \text{or} \quad k_j^2(z)=0$. 
Table \ref{CD} summarizes the results for $\mathrm{C}^g$ and $\mathrm{D}^g$ terms which are needed to carry on computations.
\begin{table}
$$
\begin{array}{|c|c|c|}\hline
&    e^\mu = \frac{1}{2} \AL{i}\g^\mu \SR{j} & e^\mu = \frac{1}{2} \AL{j}\g^\mu \SR{i} 
\\ 
\hline
&  (h_i, h_j) = (+,+) 
&  (h_i, h_j) = (-,-)
\\
\mathrm{C}^g 
& \SR{\hat{k}_i} = \sqrt{x_i}\, \SR{i} \, , \,\,\, \AR{\hat{k}_i} = \frac{\slashk_i\SR{j}}{\sqrt{x_i}\SSS{ij}} 
& \AR{\hat{k}_i} = \sqrt{x_i}\, \AR{i} \, , \,\,\, \SR{\hat{k}_i} = \frac{\slashk_i\AR{j}}{\sqrt{x_i}\AA{ij}}
\\ 
& \kstrhat_j =   \frac{\AL{j} \slashk_i + \slashk_j \SR{i}}{\SSS{ji}}  \,\,\, \text{or} \,\,\, \AR{\hat{j}} = \frac{\left( \slashk_i + \slashp_j \right)\SR{i}}{\SSS{ji}} 
& \kapphat_j = \frac{\AL{i} \slashk_i + \slashk_j \SR{j}}{\AA{ji}}  \,\,\, \text{or} \,\,\, \SR{\hat{j}} = \frac{\left( \slashk_i + \slashp_j \right)\AR{i}}{\AA{ji}}
\\
\hline
&  (h_i, h_j) = (-,-) 
&  (h_i, h_j) = (+,+)
\\
\mathrm{D}^g 
& \AR{\hat{k}_j} = \sqrt{x_j}\, \AR{j} \, , \,\,\, \SR{\hat{k}_j} = \frac{\slashk_j\AR{i}}{\sqrt{x_j}\AA{ji}}
& \SR{\hat{k}_j} = \sqrt{x_j}\, \SR{j} \, , \,\,\, \AR{\hat{k}_j} = \frac{\slashk_j\SR{i}}{\sqrt{x_j}\SSS{ji}} 
\\ 
& \kapphat_i =  \frac{\AL{j} \slashk_j + \slashk_i \SR{i}}{\AA{ij}}  \,\,\, \text{or} \,\,\, \SR{\hat{i}} = \frac{\left( \slashk_j + \slashp_i \right)\AR{i}}{\AA{ij}}
& \kstrhat_i   =  \frac{\AL{i} \slashk_j + \slashk_i \SR{j}}{\SSS{ij}}  \,\,\, \text{or} \,\,\, \AR{\hat{i}} = \frac{\left( \slashk_j + \slashp_i \right)\SR{i}}{\SSS{ij}}  
\\
\hline
\end{array}
$$
\caption{
Shifted quantities needed for the evaluation of $\mathrm{C}^g$ and $\mathrm{D}^g$ residues.
The reported helicity of the on-shell gluon is meant in the case it is on-shell. }
\label{CD}
\end{table} 
%

\begin{align}
\mathrm{A}^f_{p,h} &= \graph{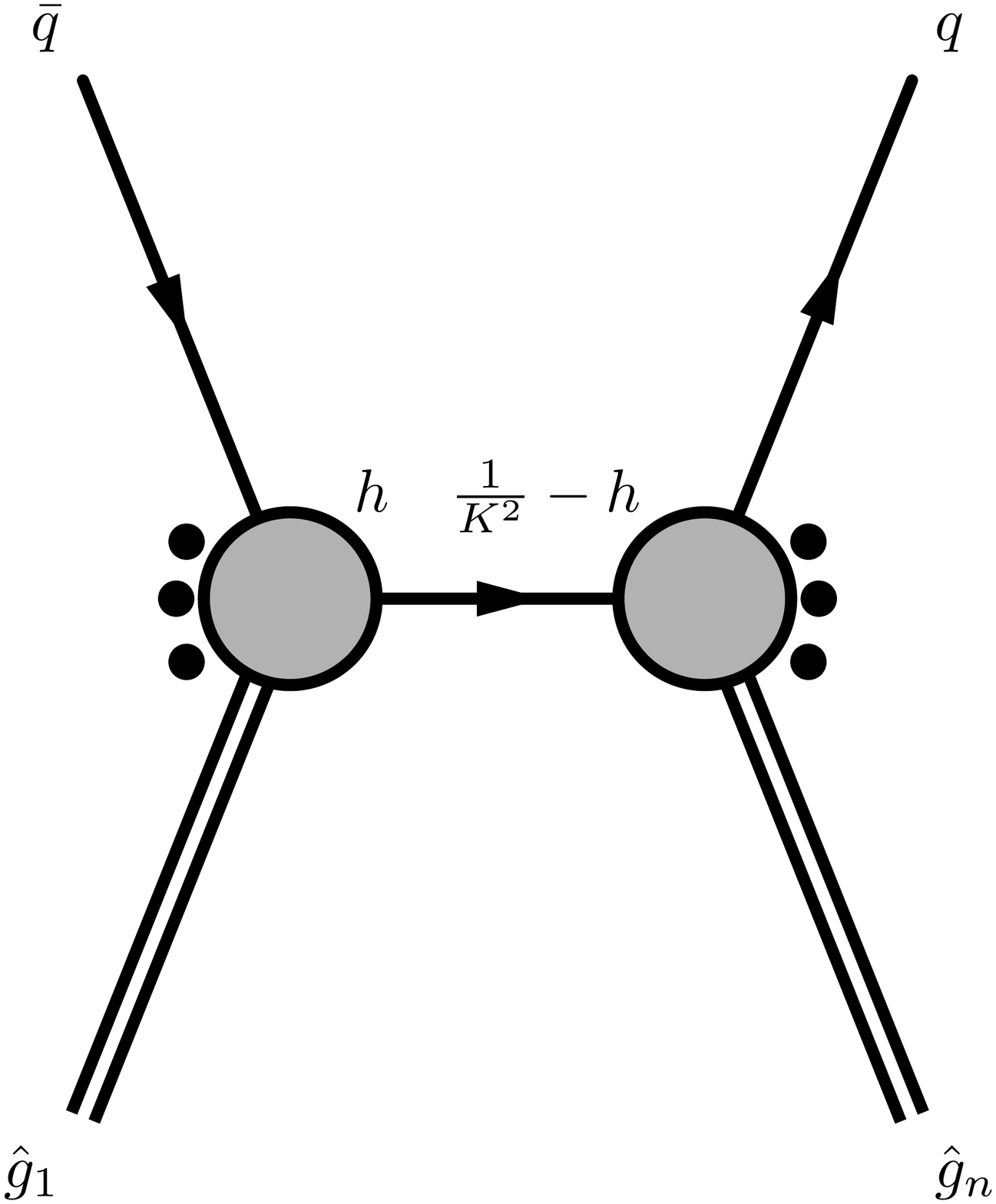}{20}{12}
\quad&
\mathrm{B}^f_{i} &= \graph{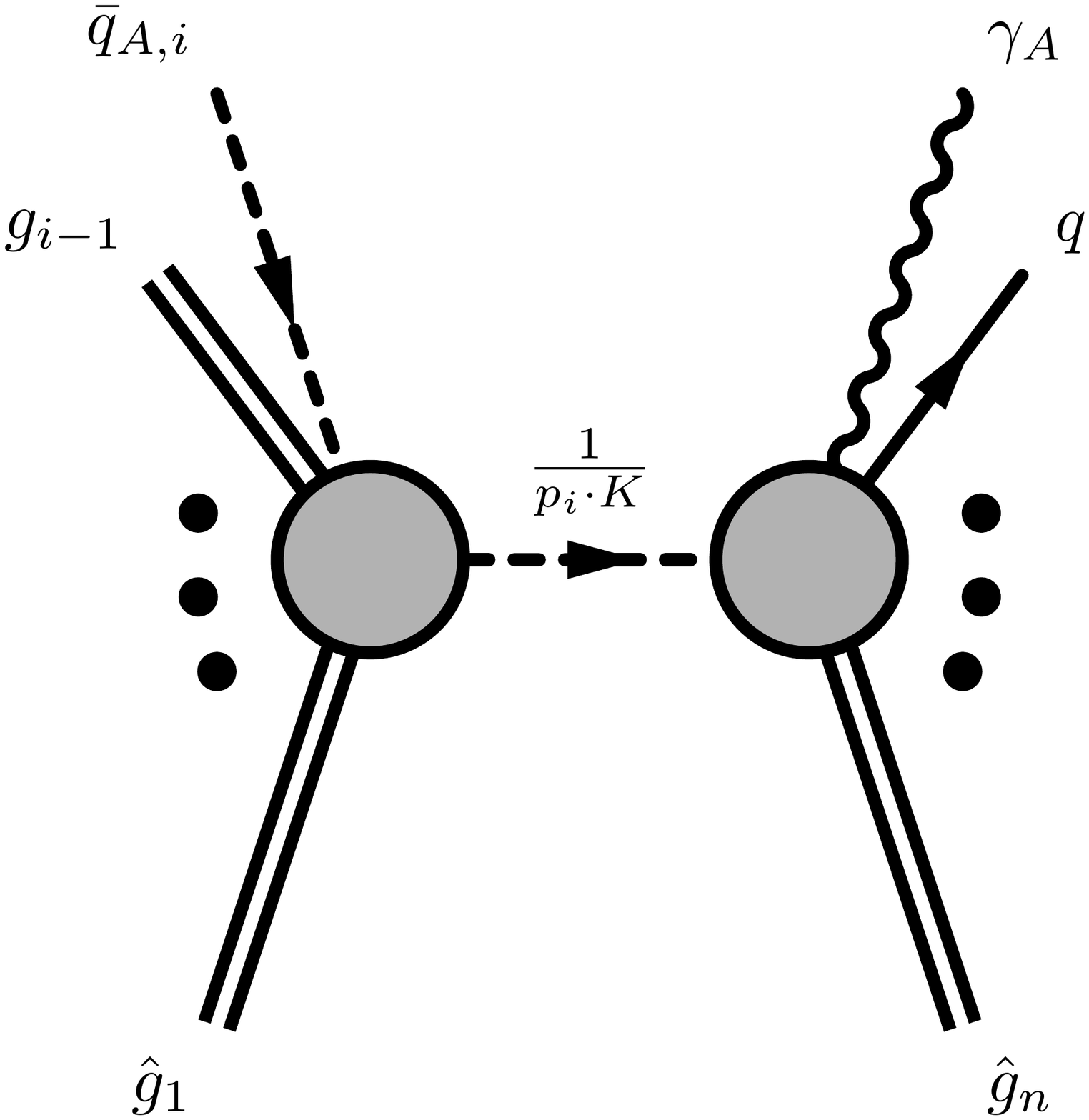}{20}{12}
\notag\\
\mathrm{C}^f &= \graph{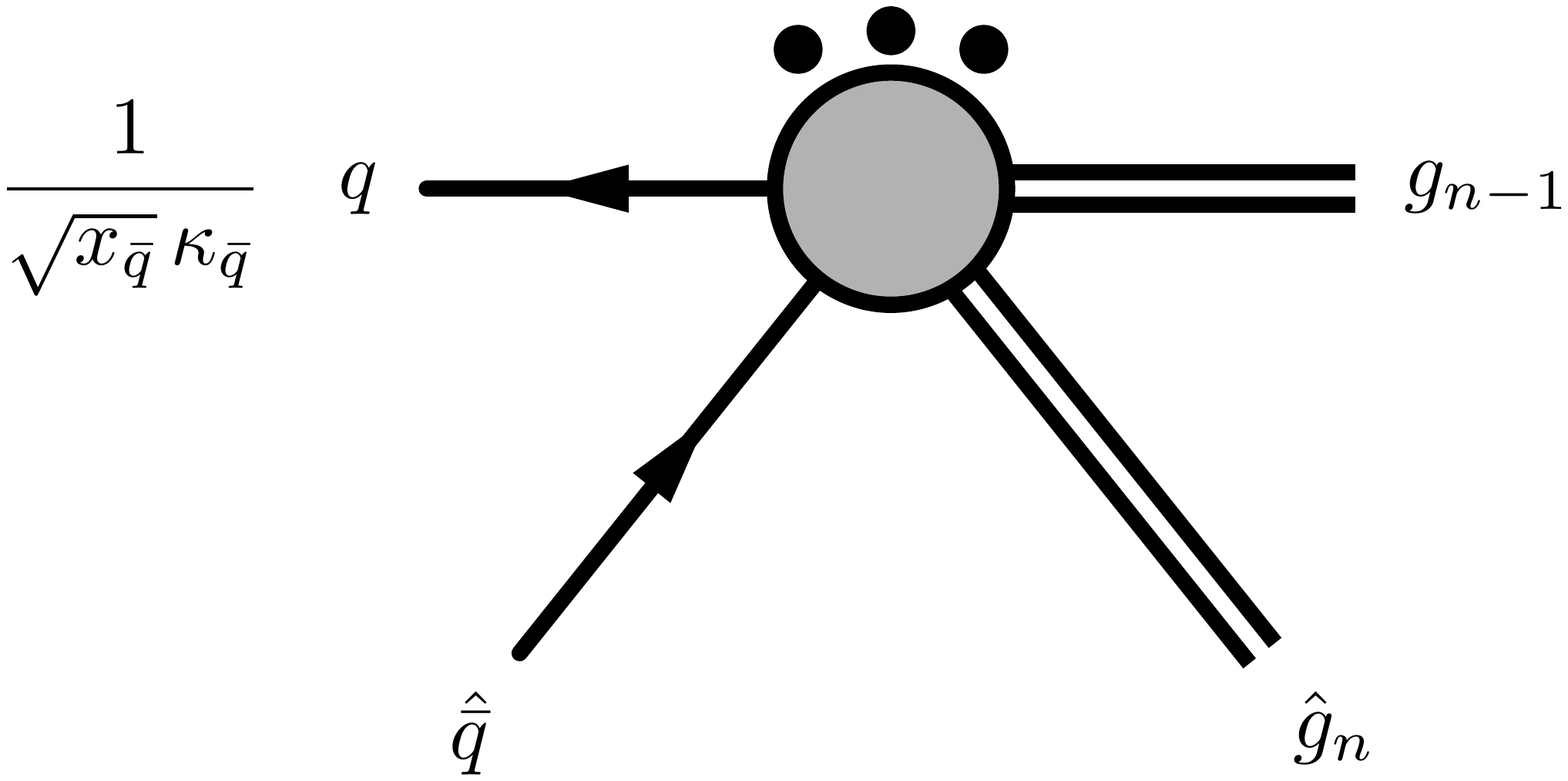}{25}{20}
\quad&
\mathrm{D}^f &= \graph{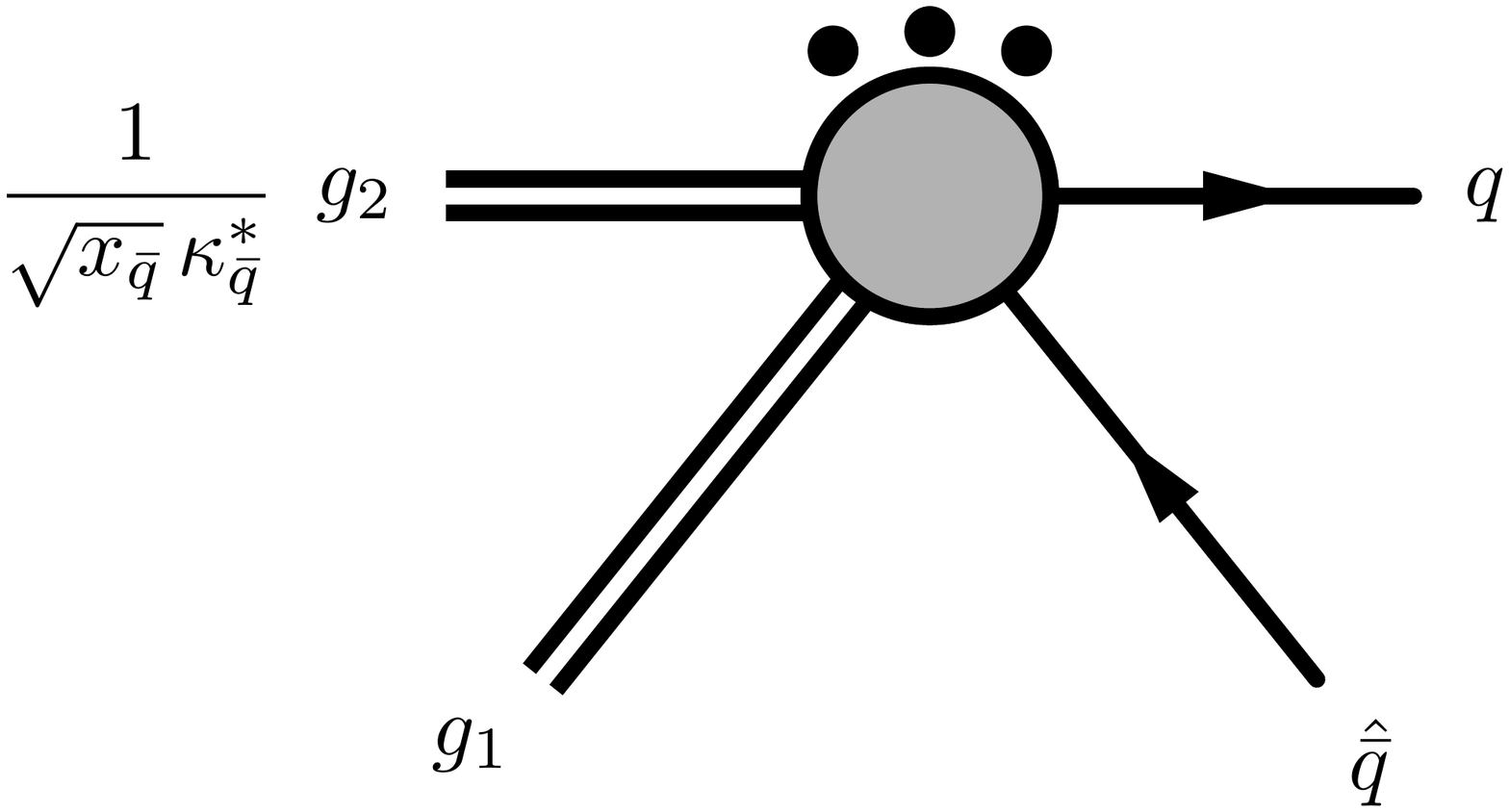}{25}{20}
\nn
\end{align}

The terms $\mathrm{A}^f_{p,h}$ and $\mathrm{B}^f_i$ are exactly the same as in the gluon case.
The denominators of propagators behave in the same way and the numerator in the case of the $\mathrm{A}^f_{p,h}$ terms ensures 
that there are opposite helicities at the ends of the propagator.
In the $\mathrm{B}^f_i$ terms we have an eikonal (anti-)quark and a zero-momentum photon needed to account for the off shell (anti-)quark
If the $i$-th particle is an on-shell fermion, they are not present.

$\mathrm{C}^f$ and $\mathrm{D}^f$ denote exactly the same kind of residues as in the case of gluons.
We summarize the results for $\mathrm{C}^f$ and $\mathrm{D}^f$ terms assuming we shift the antiquark. 
Shifting the quark is exactly the same, with obvious changes of labels.

The $\mathrm{C}^f$ and $\mathrm{D}^f$ terms identically vanish if the shift vectors 
are chosen in such a way that in the on-shell limit they reduce to a legitimate choice.
All the results needed to work out these $C^f$ and $D^f$-residues are listed in Table \ref{CDf}.
\begin{table}
$$
\begin{array}{|c|c|c|}\hline
&   \text{Admitted on-shell} & \text{Not admitted on-shell }
\\ 
\hline
&  (h_{\qb}, h_g) = (-,+)  &  (h_{\qb}, h_g) = (+,+)
\\
\mathrm{C}^f
& \SR{\hat{k}_{\qb}} = \sqrt{x_{\qb}}\, \SR{\qb} \, , \,\,\, \AR{\hat{k}_{\qb}} = \frac{\slashk_{\qb} \SR{g}}{\sqrt{x_{\qb}} \SSS{\qb g}}  
& \SR{\hat{k}_{\qb}} = \sqrt{x_{\qb}}\, \SR{\qb} \, , \,\,\, \AR{\hat{k}_{\qb}} = \frac{\slashk_{\qb} \SR{g}}{\sqrt{x_{\qb}} \SSS{\qb g}} 
\\ 
e^\mu = \frac{\AL{\qb}\g^\mu \SR{g}}{2} 
& \kstrhat_g =   \frac{\AL{g} \slashk_{\qb} + \slashk_g \SR{\qb}}{\SSS{g \qb}}  \,\,\, \text{or} \,\,\, \AR{\hat{g}} = \frac{\left( \slashk_{\qb} + \slashp_g \right)\SR{{\qb}}}{\SSS{g \qb}} 
& \kstrhat_g =   \frac{\AL{g} \slashk_{\qb} + \slashk_g \SR{\qb}}{\SSS{g \qb}}  \,\,\, \text{or} \,\,\, \AR{\hat{g}} = \frac{\left( \slashk_{\qb} + \slashp_g \right)\SR{{\qb}}}{\SSS{g \qb}}
\\
\hline
&  (h_g, h_{\qb}) = (-,+) &  (h_g, h_{\qb}) = (-,-)
\\
\mathrm{D}^f
& \AR{\hat{k}_{\qb}} = \sqrt{x_{\qb}}\, \AR{\qb} \, , \,\,\, \SR{\hat{k}_{\qb}} = \frac{\slashk_{\qb} \AR{g}}{\sqrt{x_j}\AA{\qb g}}
& \AR{\hat{k}_{\qb}} = \sqrt{x_{\qb}}\, \AR{\qb} \, , \,\,\, \SR{\hat{k}_{\qb}} = \frac{\slashk_{\qb} \AR{g}}{\sqrt{x_j}\AA{\qb g}}
\\ 
e^\mu = \frac{\AL{g}\g^\mu \SR{\qb}}{2}
& \kapphat_g =  \frac{\AL{\qb} \slashk_{\qb} + \slashk_g \SR{g}}{\AA{g \qb}}  \,\,\, \text{or} \,\,\, \SR{\hat{g}} = \frac{\left( \slashk_{\qb} + \slashp_g \right)\AR{g}}{\AA{g \qb}}
& \kapphat_g =  \frac{\AL{\qb} \slashk_{\qb} + \slashk_g \SR{g}}{\AA{g \qb}}  \,\,\, \text{or} \,\,\, \SR{\hat{g}} = \frac{\left( \slashk_{\qb} + \slashp_g \right)\AR{g}}{\AA{g \qb}}
\\
\hline
\end{array}
$$
\caption{
Shifted quantities needed for the evaluation of $\mathrm{C}^f$ and $\mathrm{D}^f$ residues in the case in which the antiquark is shifted.
A completely similar pattern holds when $q$ is shifted. The reported helicity of the gluon is meant in the case it is on-shell.}
\label{CDf}
\end{table} 
%

\section{Some explicit results}%

We will not report here the explicit derivations of scattering amplitudes, 
but rather stress the generally interesting features of the results.
Our results always concern amplitudes with one fermion pair and only one off-shell particle.

First, the situation with MHV amplitudes is quite different with respect to the on-shell case, 
where MHV amplitudes are always characterized by the Parke Taylor structure~\cite{Parke:1986gb},
which can easily be proved recursively.
For instance, in the mostly-plus case all the on-shell MHV amplitudes for Yang-Mills theories with fermions are given by
\bea
\Amp(g_1^+,\dots,g_i^-,\dots,g_j^-,\dots,g_n^+) &=& \frac{\AA{ij}^4}{\AA{12}\AA{23}\dots \AA{n1}} \, , \nn \\
\Amp(\qb^+,\dots, g_i^-,\dots,g_n^+,q^-) &=& \frac{\AA{i q}^3 \AA{i \qb}}{\AA{\qb 1}\AA{12}\dots\AA{nq}\AA{q1}} \, .
\label{MHVs}
\eea
The mostly-minus MHV amplitudes are just obtained by $\AA{ab} \leftrightarrow \SSS{ba}$. \\
In the off-shell case, one thing stands out with respect to this pattern: 
amplitudes with all gluons having the same helicity, one fermion particle with opposite helicity and one off-shell fermion particle do not vanish,
despite their on-shell limit is zero. For example $\Amp(\qb^*,q^-,g_1^+,\dots,g_n^+) \neq 0$.
We call these amplitudes {\em subleading} because their analysis and comparison to the proper MHV amplitudes
shows that they carry an extra factor whose absolute value is $\propto \sqrt{|k^2|}$, which vanishes
when the particle is on-shell and significantly suppresses the amplitude for small transverse momenta~\cite{vanHameren:2013csa}. \\
We choose to dub MHV the non-vanishing amplitudes featuring the maximum difference between 
the numbers of positive and negative helicity particles and, at the same time, not vanishing in the on-shell case.
This classification singles out our subleading amplitudes, which would be the ones with the highest value of this difference,
but it is specifically meant to be an intuitive extension of the on-shell case.

The structure of a subleading amplitude with an off-shell antiquark is given by
\bea
\Amp(g_1^+,g_2^+,\dots,g_{n-1}^+,\qb,q,g_n^+) 
&=&
\frac{- \AA{\qb q}^3}{\AA{12}\AA{23}\dots\AA{\qb q}\AA{q n}\AA{n1}} \, .
\eea
The case with an off-shell quark is completely analogous, of course.
Concerning MHV amplitudes, they can have either an off-shell (anti)fermion or an off-shell gluon.
Their structures stay exactly the same irrespectively of the relative position of the fermion pair and the gluons.
\bea
\Amp(g_1^+,g_2^+,\dots,g_{n-1}^+,\qb^*,q^+,g_n^-) 
&=& 
\frac{1}{\kstr_{\qb}}\, \frac{ \AA{\qb n}^3 \AA{q n} }{ \AA{12}\AA{23}\dots\AA{\qb q}\AA{qn}\AA{n1} } \, .
\nn \\
\Amp(g^*,\qb^+,q^-,g_1^+,g_2^+,\dots,g_n^+) 
&=&
\frac{1}{\kstr_g}\, \frac{ \AA{g q}^3 \AA{g\qb} }{\AA{g \qb} \AA{\qb q} \dots \AA{n-1 |n} \AA{ n g} } \, .
\eea

The second interesting point is that, even if all the MHV amplitudes with one off-shell particle that we computed reduce to MHV amplitudes
in the on-shell case, the converse is not true for 5-point functions.
Some 5-point amplitudes, like for example $\Amp(g_1^+,g_2^+,\qb^*,q^-,g_3^-)$ 
do not feature the maximum possible difference between the numbers of positive and
negative helicity particles and still reduce to MHV on-shell amplitudes. In the case of the amplitude above,
this is due to fermion helicity conservation forcing $\qb^*$ to become $\qb^+$ in the on-shell amplitude.
For details, we refer to~\cite{vanHameren:2015bba}.
%

\section{Summary}%

High Energy Factorization requires the computation of gauge-invariant scattering amplitudes
with non vanishing transverse component of the momenta of one or two of the incoming particles~\cite{Deak:2009xt,Kutak:2012rf}.
It was shown that it is possible to efficiently perform such calculations at tree level thanks 
to a generalization of the BCFW recursion relation to amplitudes featuring off-shell particles,.
This was first done in the pure Yang-Mills case~\cite{vanHameren:2014iua}, then fermions were recently included~\cite{vanHameren:2015bba}.
In particular, 5-point amplitudes with one off-shell parton are now completely known at tree-level~\cite{vanHameren:2015bba}. \\

\centerline{\bf \large Acknowledgements}%

This work was partially supported by NCN grant DEC-2013/10/E/ST2/00656.
The author also gratefully acknowledges support from the "Angelo Della Riccia" foundation.

\providecommand{\href}[2]{#2}\begingroup\raggedright\endgroup

\end{document}